\documentclass{article}

\usepackage{PRIMEarxiv}

\usepackage[utf8]{inputenc} % allow utf-8 input
\usepackage[T1]{fontenc}    % use 8-bit T1 fonts
\usepackage{hyperref}       % hyperlinks
\usepackage{url}            % simple URL typesetting
\usepackage{booktabs}       % professional-quality tables
\usepackage{amsfonts}       % blackboard math symbols
\usepackage{nicefrac}       % compact symbols for 1/2, etc.
\usepackage{microtype}      % microtypography
\usepackage{lipsum}
\usepackage{fancyhdr}       % header
\usepackage{graphicx}       % graphics
\graphicspath{{media/}}     % organize your images and other figures under media/ folder

\usepackage{here}
\usepackage{amsmath,amssymb}
\usepackage{subcaption}
\usepackage{color}
\usepackage{comment}
\usepackage{multirow}
\usepackage{rotating}

\title{Evolution of Collective AI\\Beyond Individual Optimization}

\author{
  Ryosuke Takata$^1$\author[* \quad\quad\quad\quad Yujin Tang$^2$ \quad\quad\quad\quad Yingtao Tian$^2$ \\
  \textbf{Norihiro Maruyama$^1$ \quad\quad Hiroki Kojima$^1$ \quad\quad Takashi Ikegami$^1$} \\\\
  $^1$Graduate School of Arts and Sciences \\
  The University of Tokyo \\
  Tokyo, Japan \\\\
  $^2$Sakana AI \\
  Tokyo, Japan \\\\
  \texttt{\author[*takata@sacral.c.u-tokyo.ac.jp}
}

\begin{document}
\maketitle

\begin{abstract}
This study investigates collective behaviors that emerge from a group of homogeneous individuals optimized for a specific capability. We created a group of simple, identical neural network based agents modeled after chemotaxis-driven vehicles that follow pheromone trails and examined multi-agent simulations using clones of these evolved individuals. Our results show that the evolution of individuals led to population differentiation. Surprisingly, we observed that collective fitness significantly changed during later evolutionary stages, despite maintained high individual performance and simplified neural architectures. This decline occurred when agents developed reduced sensor-motor coupling, suggesting that over-optimization of individual agents almost always lead to less effective group behavior. Our research investigates how individual differentiation can evolve through what evolutionary pathways.
\end{abstract}

% keywords can be removed

\begin{figure}[htbp]
\centering
\includegraphics[width=11cm]{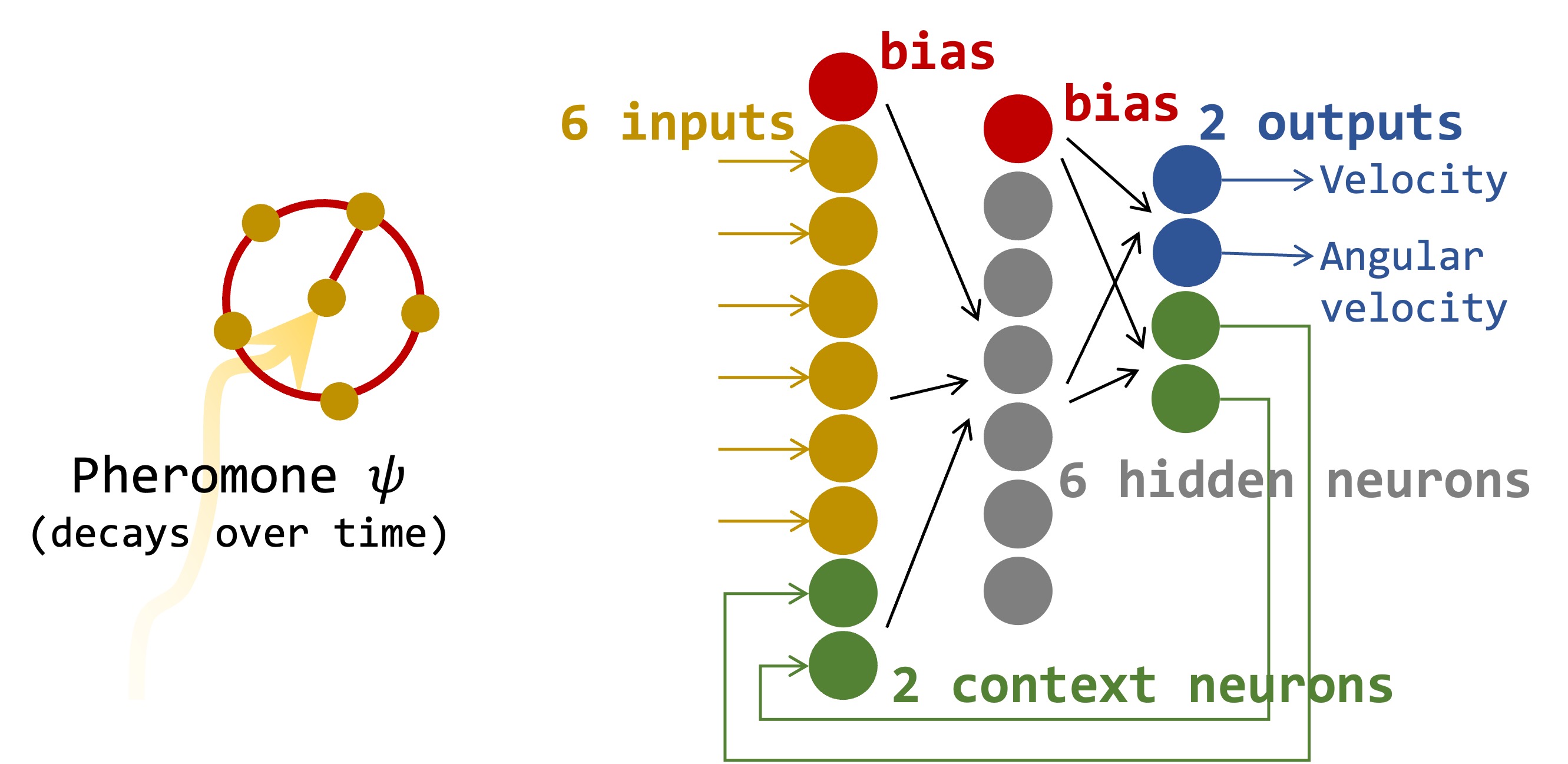}
\caption{Neural network model of an agent. The agent receives input from six sensors (five on the periphery of its circular body and one at the center) and moves using two motor outputs (linear and angular velocities). The agent's neural network consists of $6$ sensor input neurons, $6$ hidden neurons, $2$ output neurons, and $2$ context neurons, with a bias neuron in each layer.}
\label{fig:1}
\end{figure}

\section{Introduction}
Artificial Intelligence (AI) has witnessed significant advances with the emergence of powerful neural network (NN) models. Examples include large language models \cite{brown2020language} and image generation models such as DALL-E \cite{ramesh2022hierarchical}, Imagen \cite{saharia2022photorealistic}, and Parti \cite{yu2022scaling}. Each has achieved previously unseen capabilities as powerful individuals through recent technical breakthroughs.

On the other hand, the biological evolutionary strategy focuses more on the direction of collective intelligence compared to individual ability, especially for species living in populations \cite{bonabeau1999swarm}. Unlike individual intelligence, which deals with challenges independently, collective intelligence necessitates the ability to process information, operate in a decentralized manner, and adaptively integrate information based on context. This distinction is evident in social insects, such as ants and bees, where collective behavior with role differentiation emerges not from highly complex individuals but through simple interactions among members.

Recent studies have demonstrated various approaches to understanding collective behavior. Neural controllers evolved to climb environmental gradients can develop exploration-exploitation strategies \cite{van2022environment}, while effective gradient climbing can emerge from local sensing and interactions without global planning \cite{karaguzel2023collective}. The exploration processes themselves show remarkable parallels between environmental and internal memory search, suggesting common evolutionary origins \cite{todd2020foraging}. Social interactions shape exploration strategies that may not necessarily optimize group-level efficiency \cite{garg2024evolution}. Physical approaches to collective behavior in insects have revealed how individual properties and their interactions lead to various scales of collective patterns \cite{shishkov2022social}. These findings suggest that collective intelligence emerges through the complex relationship between individual and group-level behaviors \cite{watson2023collective}, where local interactions and environmental feedback play crucial roles.

We seek to bridge the \emph{biological} collective intelligence with \emph{artificial} intelligence, which we denote as ``Collective AI'' \cite{ha2022collective} and \cite{takata2023evolving}, focusing on the use of a population of neural networks.
Here the intelligence of Collective AI is embodied not in the capacity of individual neural network, but instead in how each member (or agent) of the population interacts with each other, as well as the dynamics of the resulting population. This interaction can be quantified through information-theoretic measures, particularly focusing on how environmental information is processed and integrated with internal states to generate collective behaviors.

In this study, we focus on how individual optimization of a specific behavior--chemotaxis--can lead to emergent collective behavior when agents interact in a shared environment. Specifically, we investigate the evolution of collective intelligence in populations of neural network-controlled agents inspired by clonal insects. These agents, evolved individually to perform chemotaxis, interact in multi-agent simulations where communication occurs via pheromone signals. In our simulation, we observe that collective intelligence is realized, which accompanies role differentiation. Individuals that specialize in collecting chemicals in the environment can effectively use the same chemicals for communication when they are in a group. While individual evolution prioritizes optimization for specific tasks, we seek to understand how group-level dynamics emerge and diverge from individual optimization.

\section{Models}
In the model, we first induce chemotaxis in a single agent: in the framing of chemotaxis, this agent then senses the chemicals produced by other individuals, using this chemotactic response to gather with them. The resulting collective behavior is subsequently analyzed.

\subsection{Agents Controlled by Neural Networks}
An agent's behavior is controlled by a neural network (Figure \ref{fig:1}), the parameters of which changes through an neuroevolution process. The neural network has an input layer of $9$ neurons ($1$ bias, $6$ input neurons, $2$ context neurons), a hidden layer of $7$ neurons ($1$ bias, $6$ hidden neurons), and an output layer of $4$ neurons ($2$ output neurons, $2$ context neurons). Neuroevolution was performed on all $82$ weights of the neural network. We used CMA-ES \cite{Hansen2016} in EvoJAX \cite{Yujin2022} as our evolutionary algorithm. CMA-ES generates multiple candidate solutions using a multivariate normal distribution and calculates their goodness of fit. The advantage here is that it is easy to parallelize the evaluation of the objective function with the number of candidate solutions.

The agent moves by having its velocity and angular velocity determined by a neural network. The range of values for the angular velocity $\omega(t)$ is $[-0.05, 0.05]$, and the following formula is used to update the agent's angle $\theta(t)$.

\begin{equation}
\theta(t) = \theta(t-1) + \omega(t)
\end{equation}

The range of values for velocity $v(t)$ is $[0.0, 1.0]$, and the following formula is used to update the agent's position $x(t)$ and $y(t)$. Here, the field has periodic boundary conditions.

\begin{equation}
x(t) = x(t-1) + v(t) \cdot \cos(\theta(t))
\end{equation}
\begin{equation}
y(t) = y(t-1) + v(t) \cdot \sin(\theta(t))
\end{equation}

\begin{comment}
\begin{figure}[htbp]
\centering
\includegraphics[width=10cm]{fig/01.jpg}
\caption{Neural network model of an agent. The agent receives input from six sensors (five on the periphery of its circular body and one at the center) and moves using two motor outputs (linear and angular velocities). The agent's neural network consists of $6$ sensor input neurons, $6$ hidden neurons, $2$ output neurons, and $2$ context neurons, with a bias neuron in each layer.}
\label{fig:1}
\end{figure}
\end{comment}

\subsection{Experimental Setup}
The experiment consists of two phases: (1) an evolution phase where a single agent's neural network is trained to acquire chemotaxis behavior, and (2) a test phase where the evolved neural network is replicated across $1024$ agents for evaluation. In the test phase, evaluation is conducted on a homogeneous population where all agents share identical neural network parameters.

We evaluate our evolved neural network periodically. Concretely, for every $10$ generations, we pick the best individual from the population and replicate its parameters (i.e. the synaptic weights) $1024$ times to create identical agents, and use these agents to conduct a multi-agent simulation.

The parameter values used in the simulations are detailed in Table \ref{tab:1}. The environmental parameters, such as spatial dimensions and pheromone evaporation rate, remain identical between the single-agent evolution phase and the multi-agent test phase.

\begin{table}[htbp]
\small\sf\centering
\caption{Summary of the simulation parameters.\label{tab:1}}
\begin{tabular}{ll}
\toprule
Parameter&Value\\
\midrule
Field width & 600\\
Field height & 600\\
Agent body radius & 20\\
Agent sensor radius & 2\\
Number of sensors on agent & 6\\
Number of agent actions & 2\\
Number of initial pheromone spots & 5\\
Pheromone decay rate & 0.001\\
Maximum number of steps (for a single agent) & 1000\\
Maximum number of steps (for multi agent) & 5000\\
Maximum generations & 2000\\
Population & 100\\
Initial standard deviation of candidate solution individuals & 0.1\\
\bottomrule
\end{tabular}
\end{table}

\subsection{Pheromone Environment}
\begin{figure}[htbp]
\centering
\includegraphics[width=10cm]{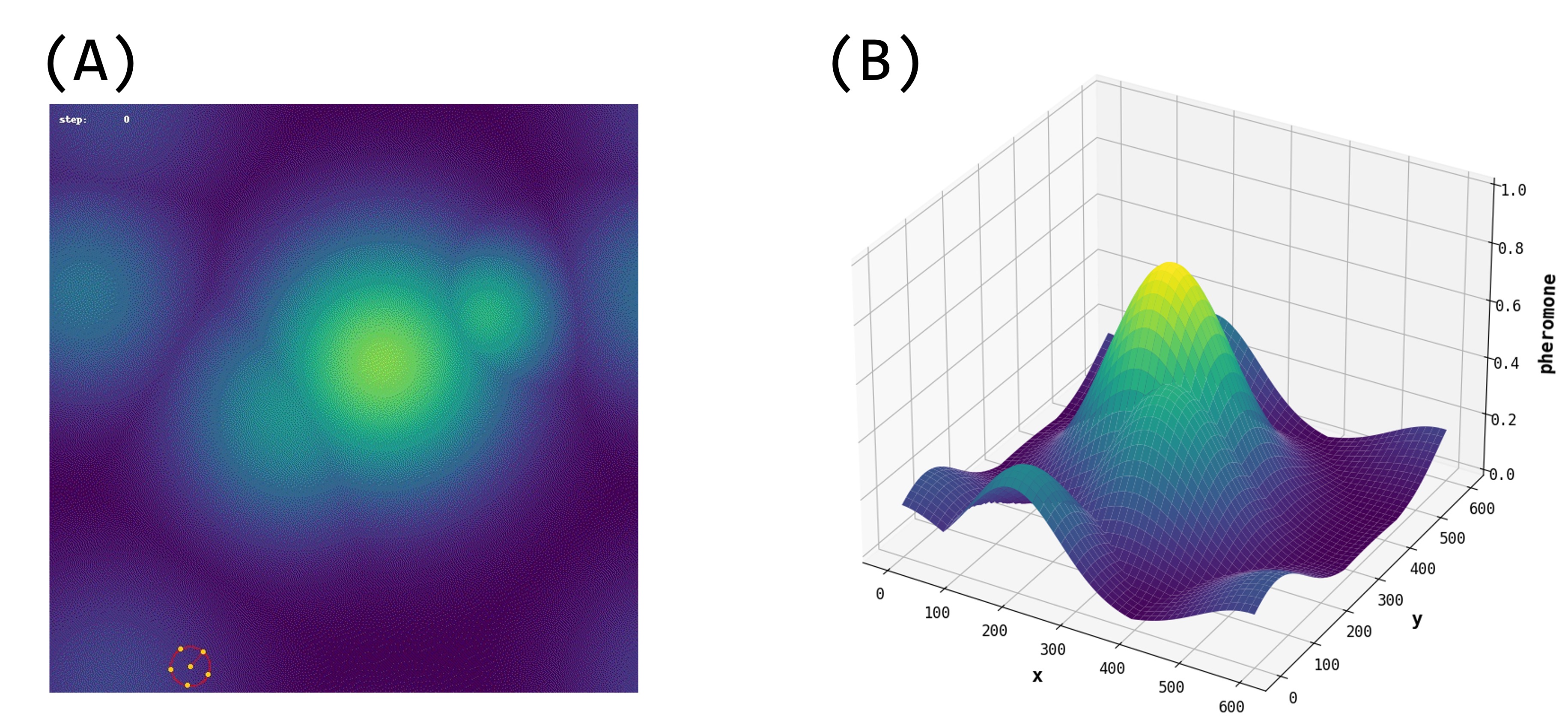}
\caption{Single agent evolutionary simulation environment.
\textbf{(A)} 2D plot of pheromones and the agent. The initial placement of both the pheromones and the agent is randomized for each trial.
\textbf{(B)} 3D plot of pheromones. These pheromones follow a gradient based on a mixed Gaussian distribution, decaying by $0.001$ per step.}
\label{fig:2}
\end{figure}

The evolution phase which agents are subjected to chemotaxis, the environment is populated with a single agent and pheromones (Figure \ref{fig:2}). The pheromones evaporate and decay over time. In this phase, Agents can evolve to gain the fitness. The fitness function is expressed by the following function:

\begin{equation}
F(t) = \int \psi(\vec{r}(t), t) dt
\end{equation}
where $\psi$ is the amount of pheromone released into the environment other than its own release, and $\vec{r}(t)$ is the position of the agent at time $t$.

In the evolution phase, the placement of the pheromones changes randomly with every trial. Five initial pheromones are generated at random locations. The shape of the initial pheromone is a mixed bell-shaped distribution with randomly assigned parameters. One bell-shaped distribution has the parameters maximum amount of pheromone $a$, standard deviation $\sigma$, and center coordinates $x_c$, $y_c$, and is expressed by the following probability density function. Here, Table \ref{tab:2} summarizes the details of each parameter that is randomly set.

\begin{equation}
f(x, y) = a \cdot \exp{\bigg\{-\frac{(x - x_c)^2 + (y - y_c)^2}{2 \sigma^2}\bigg\}}
\end{equation}

In the multi-agent simulation, instead of placing initial random pheromones, each of the $1024$ agents releases its own pheromones. Each agent deposits pheromones with a fixed value of $1.0$ in a $3 \times 3$ area centered on its previous position, overwriting any existing pheromones in that area. The environment contains only one type of pheromone, and pheromones in the environment decay at a rate of $0.001$ per step. This mechanism allows agents to communicate with each other through applying changes to their environment. In this test phase, how much pheromones collected by the agents are computed as the measure of collective performance.

\begin{table}[htbp]
\small\sf\centering
\caption{Summary of value ranges for initial pheromone shape parameters. \label{tab:2}}
\begin{tabular}{ll}
\toprule
Parameter&Range\\
\midrule
Maximum amount of pheromone $a$ & $[0.2, 1.0]$\\
Standard deviation $\sigma$ & $[50, 100]$\\
Center coordinate $x_c$ & $[2, 597]$\\
Center coordinate $y_c$ & $[2, 597]$\\
\bottomrule
\end{tabular}
\end{table}

\section{Results}
\subsection{Local vs. Global Movement Patterns} 
\begin{figure}[htbp]
\centering
\includegraphics[width=\columnwidth]{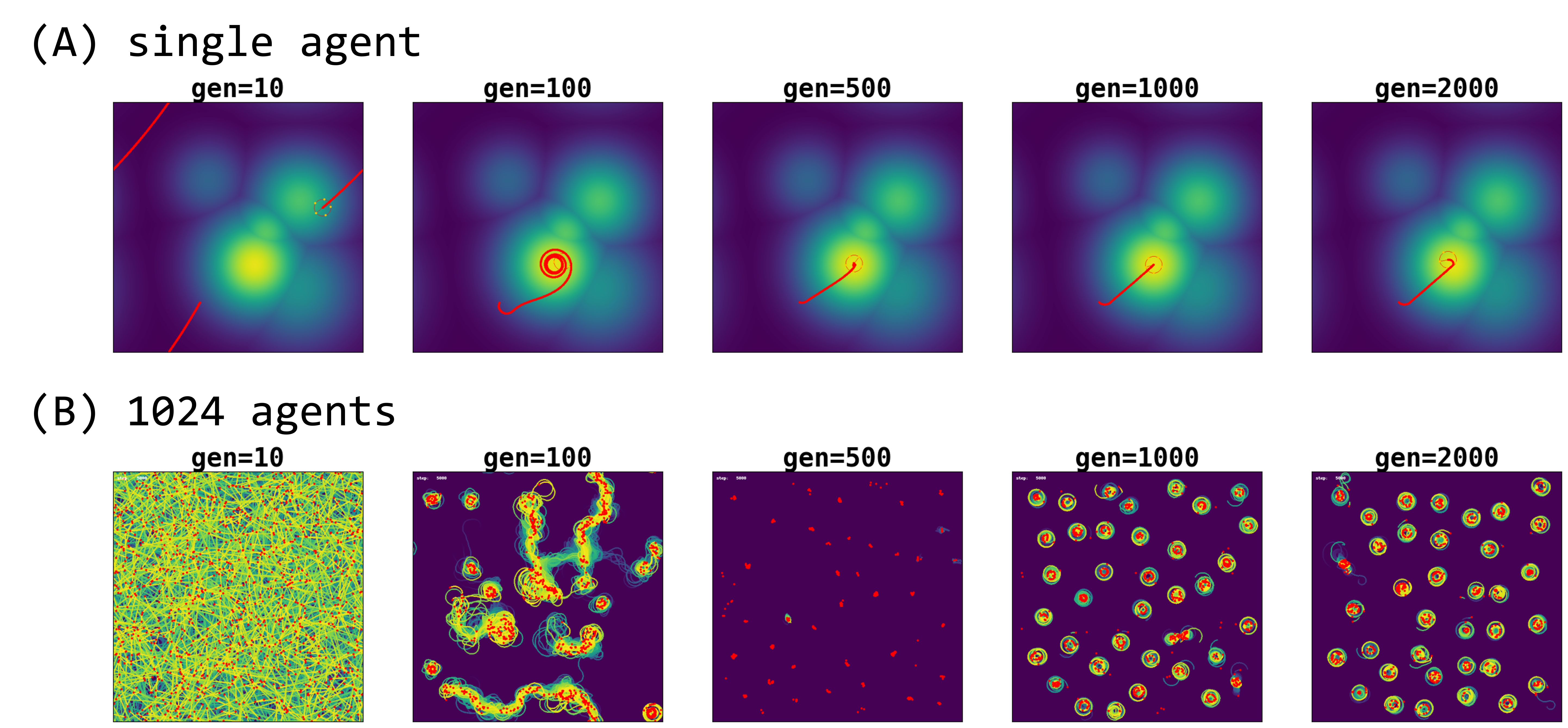}
\caption{Representative agent trajectories at generations $10$, $100$, $500$, $1000$, and $2000$.
\textbf{(A)} Trajectories of an evolved single agent in a test environment. The red line represents the agent's path over $1000$ steps. The evolution shows that the agent has acquired chemotaxis.
\textbf{(B)} Snapshot at $5000$ steps in a multi-agent simulation with $1024$ agents. Each agent releases pheromones, and each red dot marks the central position of an agent. Three distinct collective behaviors emerge: (1) field-wide spreading, (2) formation of wavy lines, and (3) localized gathering.}
\label{fig:3}
\end{figure}

The evolution of agents' movement patterns was compared between the single agent and multi-agent simulations (Figure \ref{fig:3}). In the first generation, both in the single agent and in multi-agent, they do not respond to pheromones and move only linearly. This is because the neural network has not yet evolved to properly respond to sensor inputs and change motor outputs.

As the generations progressed, the single agent acquired chemotaxis, which also changed the overall behavior of the multi-agent. At generation $100$, the single agent does not optimize its behavior to stop at the position of maximum pheromone concentration, and multi-agent simulations differentiate agent movement patterns. Within the population of agents sharing the same neural network, some exhibit local movement patterns while others roam more globally. Even when agents settle into localized patterns, disrupting these local structures can transition them into more global movements.

Then at generation $500$, the single agent shows an optimized movement pattern, and in multi-agent simulations, agents cluster locally at multiple positions. In these clusters, most agents achieve the optimal strategy of continuously gaining pheromones by remaining stationary, similar to the single agent case. After this point, while the single agent's movement pattern remains largely unchanged, the multi-agent simulation shows further evolution: after generation $1000$, the spatial size of local clusters slightly increases, forming dynamic clusters that continue to move. This results in a decrease in the collective fitness measured by pheromone gain, which will be discussed later.

We also analyzed the time series of pheromone gain in this environment (see Figures \ref{fig:a1} and \ref{fig:a2} in the Appendix). The analysis showed that single agents evolved chemotaxis behavior that enables them to reach areas of highest pheromone concentration via optimal paths.

\subsection{Phase Transition of Collective Behavior}
\begin{figure}[htbp]
\centering
\includegraphics[width=\linewidth]{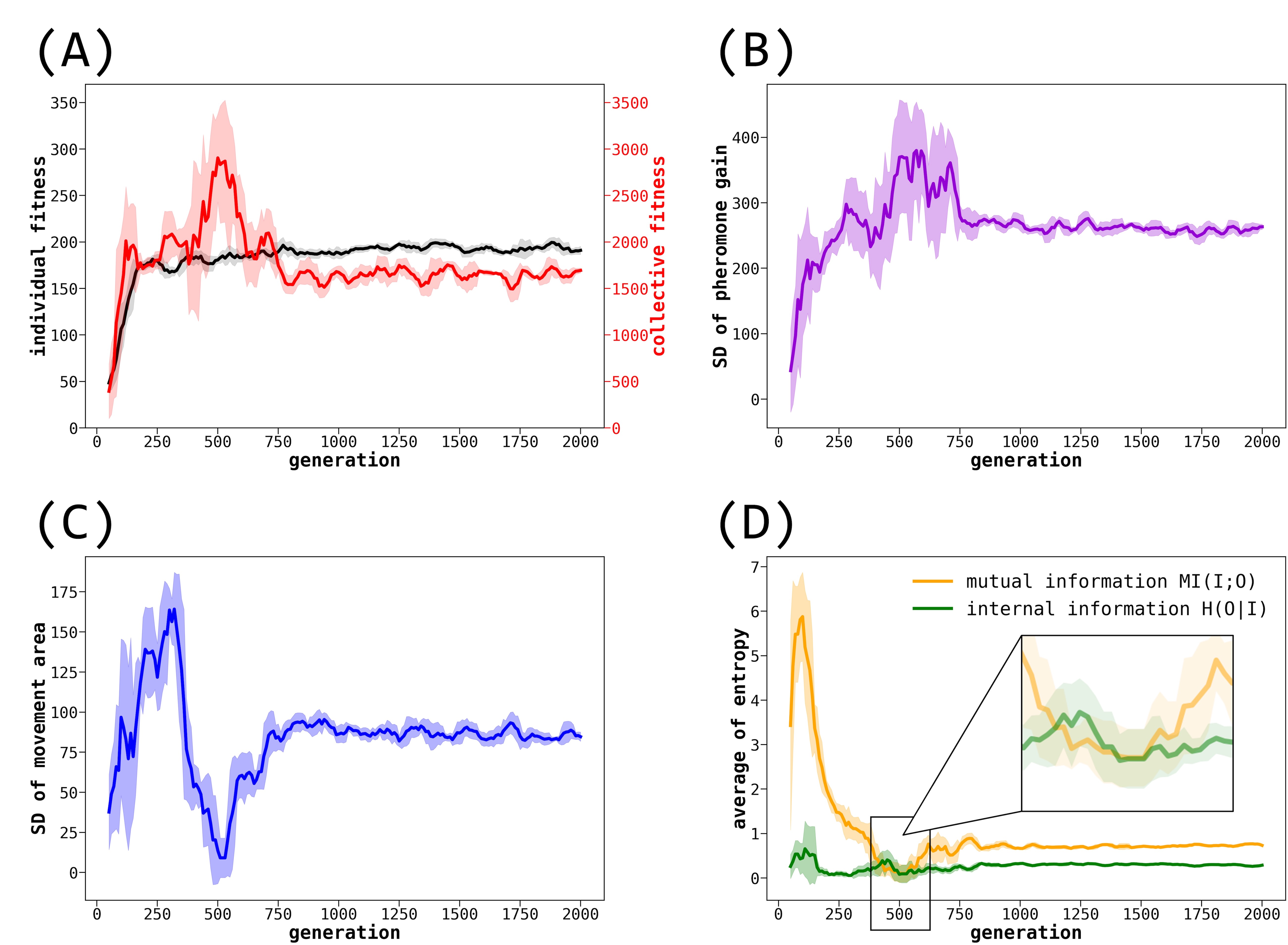}
\caption{Moving averages of statistical metrics for $1024$ agents across generations for a representative evolutionary seed.
\textbf{(A)} Collective fitness shows significant variation after individual fitness converges. In the middle generations, collective fitness increases, then declines, and eventually stabilizes.
\textbf{(B)} Standard deviation of pheromone gain among $1024$ agents. The pattern mirrors the transitions in collective fitness (A), suggesting that higher collective fitness correlates with greater differentiation in pheromone gain.
\textbf{(C)} Standard deviation of movement areas among $1024$ agents. During generations with high collective fitness (A), despite greater differentiation in pheromone gain (B), agents show similar movement patterns between individuals.
\textbf{(D)} Average of mutual information between sensor inputs and motor outputs $MI(\boldsymbol{I};\boldsymbol{O})$ and contribution of internal states to outputs $H(\boldsymbol{O}|\boldsymbol{I})$. It shows that the sensor-motor coupling has decreased throughout evolution. Overall, there is more information from the environment than from the internal state.}
\label{fig:4}
\end{figure}

We analyzed the evolution of collective behavior (Figure \ref{fig:4}). The collective fitness (average of individual fitness across agents) showed significant variation even after individual fitness (ability to collect pheromones) converged (Figure \ref{fig:4} (A)). This analysis reveals that behaviors which maximize individual benefits may not necessarily optimize the collective performance, yet group-level fitness remains quantifiable and meaningful. The multi-agent environment creates situations not present in single-agent scenarios, leading to differences in fitness diversity.

Changes in collective fitness can be explained by the differentiation of pheromone gain within the population (Figure \ref{fig:4} (B)). Groups with high collective fitness show clear separation between agents that collect few pheromones and those that collect many.

However, groups with high collective fitness show similar movement patterns across agents (Figure \ref{fig:4} (C)). This indicates that despite similar movement patterns, there is variation in pheromone gain. In the localized cluster patterns seen at generation $500$ (Figure \ref{fig:3} (B)), while agents exhibit nearly identical stationary behavior patterns, their pheromone gain differs due to differentiation between large and small clusters.

To analyze the contributions of different information sources to agent behavior, we examined both external information from sensor inputs and internal state information including context neurons in the neural network. We calculated the mutual information $MI(\boldsymbol{I};\boldsymbol{O})$ between sensor inputs and motor outputs, and the conditional entropy $H(\boldsymbol{O}|\boldsymbol{I})$ of outputs conditioned on sensor inputs (Figure \ref{fig:4} (D)). The mutual information and conditional entropy are given by the following equations:
\begin{equation}
MI(\boldsymbol{I};\boldsymbol{O}) = H(\boldsymbol{I}) + H(\boldsymbol{O}) - H(\boldsymbol{I},\boldsymbol{O})
\end{equation}
\begin{equation}
H(\boldsymbol{O}|\boldsymbol{I}) = H(\boldsymbol{I},\boldsymbol{O}) - H(\boldsymbol{I})
\end{equation}
where $\boldsymbol{I}$ represents the time series data from $6$ sensor inputs and $\boldsymbol{O}$ represents $2$ motor outputs. These calculations used sensor and motor values (min=$0$ and max=$1$) discretized with a bin width of $0.01$. And the entropy of agent behavior (motor outputs) $H(\boldsymbol{O})$ is expressed as the sum of $MI(\boldsymbol{I};\boldsymbol{O})$ and $H(\boldsymbol{O}|\boldsymbol{I})$:
\begin{equation}
H(\boldsymbol{O}) = MI(\boldsymbol{I};\boldsymbol{O}) + H(\boldsymbol{O}|\boldsymbol{I}).
\end{equation}

Overall, the mutual information is higher than the internal information content, suggesting that information from the environment significantly contributes to agent behavior. The mutual information initially increases before sharply decreasing (Figure \ref{fig:4} (D)). At generation $500$, when collective fitness reaches its peak, mutual information is at its lowest. This suggests that collective behavior becomes optimal when agents rely less on environmental information. Notably, at generation $100$, where agents display interesting ant-like patterns (Figure \ref{fig:3} (B)), both mutual information and internal information reach their highest values, with $MI(\boldsymbol{I};\boldsymbol{O})$ significantly dominating over $H(\boldsymbol{O}|\boldsymbol{I})$. This indicates that the diversity in collective behaviors emerges primarily through strong environmental coupling rather than internal state dynamics. With respect to the internal states, it can be observed in agents' context neurons, which show the most diverse patterns at generation $100$ (see Figure \ref{fig:a3} in Appendix).

\subsection{Evolution of Energy Distribution}
We still need to quantify the population dynamics. The activity of agents can be measured as kinetic energy. We measured the kinetic energy of individual $i$ at time step $t$ by tracking the agent's behavior, defined by the following equation:

\begin{equation}
E_i(t) = \Delta x_i^2 + \Delta y_i^2.
\end{equation}

\begin{figure}[htbp]
\centering
\includegraphics[width=8cm]{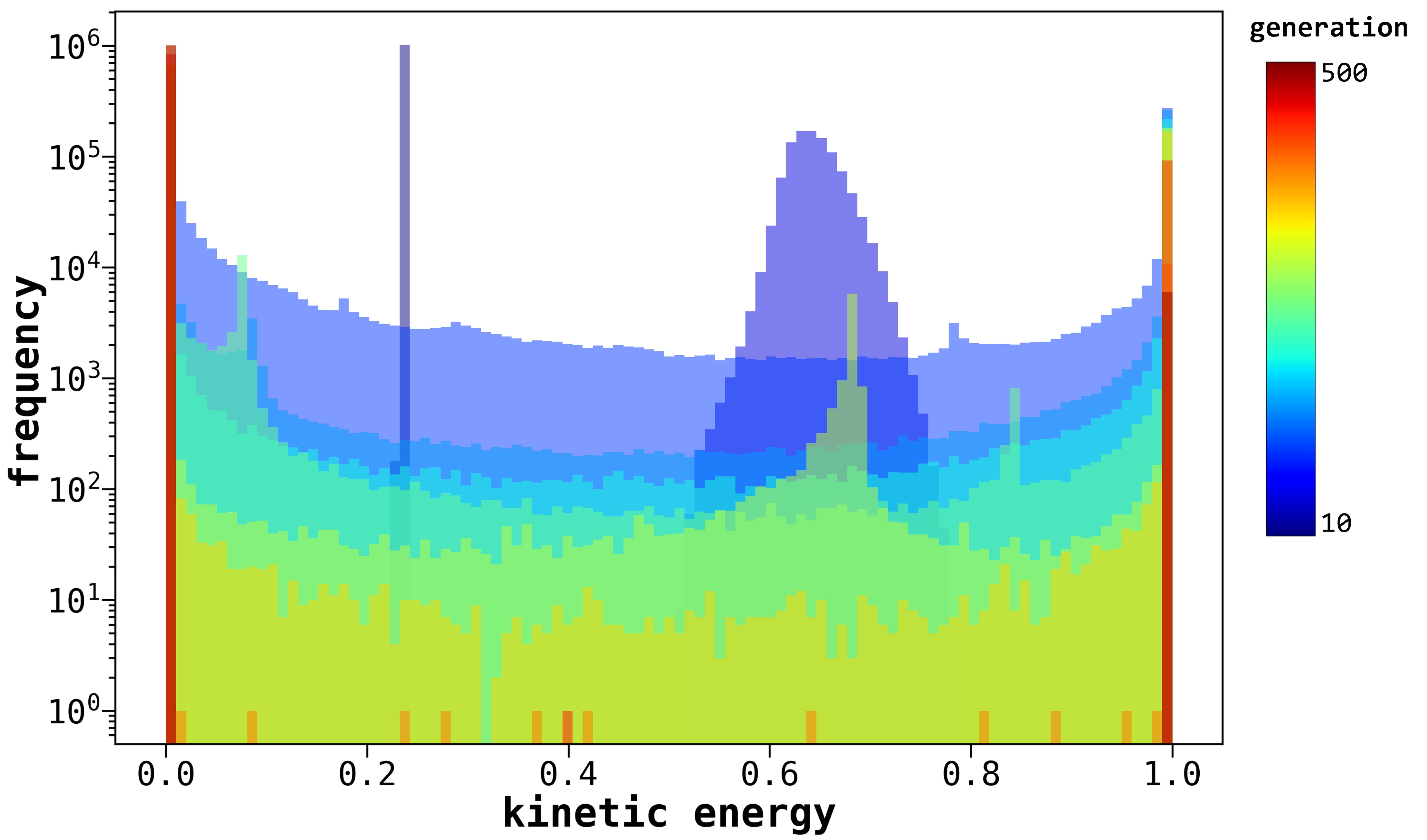}
\caption{Kinetic energy distribution among $1024$ agents up to generation $500$, calculated over $1000$ sampled steps per generation. The colors of the distributions represent different generations. As evolution progresses, the distribution becomes bimodal with values concentrated at $0$ and $1$. After generation $500$, the shape of the distribution remains largely unchanged.}
\label{fig:5}
\end{figure}

We analyzed the changes of collective motion through evolutionary dynamics by looking at their kinetic energy distributions (Figure \ref{fig:5}). It is noteworthy that the energy dispersion of agents is small in the first generation, then the dispersion increases, and in subsequent generations the energy is biased toward a maximum or minimum. In other words, as the evolution progresses, the kinetic energy tends to become more discrete, with the agents showing two extreme behaviors: either halting or moving at maximum speed. This indicates that there is diversity in behavior among agents around the generation $100$, when the mutual information between sensor inputs and motor outputs $MI(\boldsymbol{I};\boldsymbol{O})$ (Figure \ref{fig:4} (D)) is at its maximum. Conversely, by generation $500$, the heterogeneity of individual dynamics diminishes, and the mutual information between sensor inputs and motor outputs ($MI(\boldsymbol{I};\boldsymbol{O})$) approaches the entropy of internal states to outputs ($H(\boldsymbol{O}|\boldsymbol{I})$).

\subsection{Diversity in Collective Behavior}
\begin{figure}
\centering
\includegraphics[width=\linewidth]{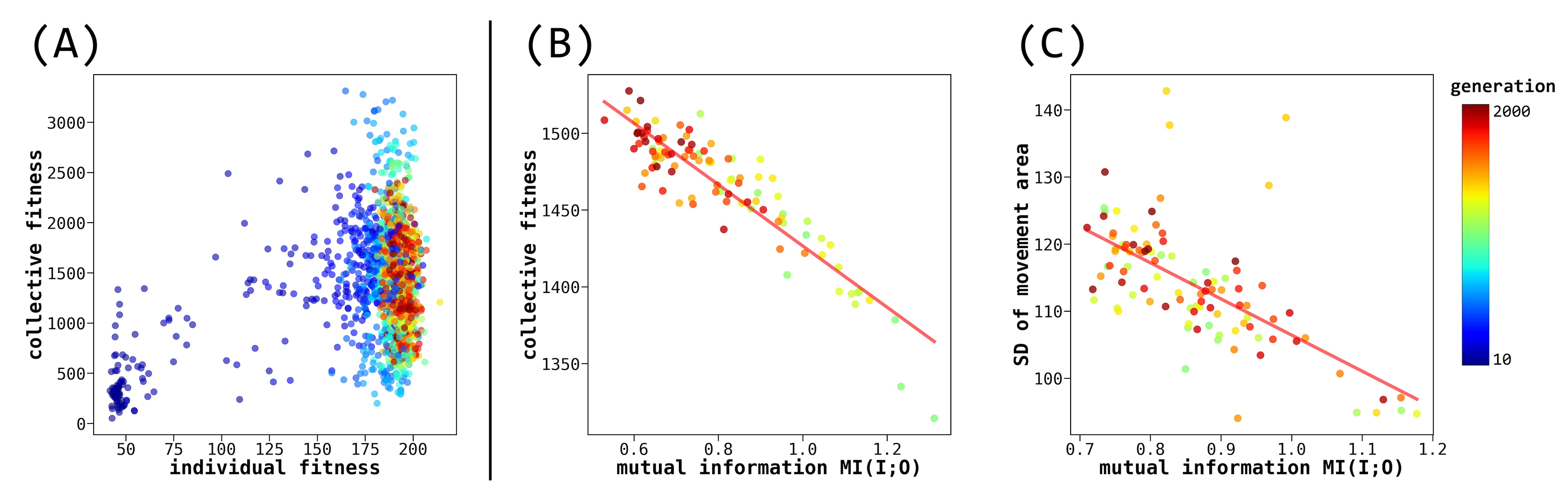}
\caption{Results from evolutionary simulations conducted with $10$ different random seeds. The colors of the plot points indicate the generations.
\textbf{(A)} Relationship between individual and collective fitness. A certain level of individual fitness leads to a diversity of collective fitness.
\textbf{(B)} Negative correlation between mutual information $MI(\boldsymbol{I};\boldsymbol{O})$ and collective fitness, shown for a representative seed from generation $1000$ onward.
\textbf{(C)} Negative correlation between mutual information $MI(\boldsymbol{I};\boldsymbol{O})$ and the standard deviation of movement areas among $1024$ agents, shown for a representative seed from generation $1000$ onward.}
\label{fig:6}
\end{figure}

We examined the relationship between individual and collective fitness across $10$ different evolutionary seeds (Figure \ref{fig:6} (A)). The results show that once individual fitness exceeds a certain level, collective fitness diversifies. This suggests that collective fitness is not uniquely determined by individual fitness.

Particularly in the later generations where collective fitness shows diversification, we investigated the relationship between mutual information $MI(\boldsymbol{I};\boldsymbol{O})$ with both collective fitness and differentiation in movement patterns. We see negative correlations in both cases (Figure \ref{fig:6} (B) and (C)). Across the $10$ seeds, the correlation coefficients between mutual information $MI(\boldsymbol{I};\boldsymbol{O})$ and collective fitness ranged from $-0.91$ to $0.03$ ($\mu=-0.41$, $s=0.26$), while those between mutual information $MI(\boldsymbol{I};\boldsymbol{O})$ and the standard deviation of movement patterns ranged from $-0.64$ to $0.21$ ($\mu=-0.21$, $s=0.29$) (Figure \ref{fig:a5}). These results indicate that lower mutual information $MI(\boldsymbol{I};\boldsymbol{O})$ tends to correlate with higher collective fitness and more uniform but rather spatially localized movement patterns.

\section{Discussions}
We demonstrated the emergence of ant-like swarms (Figure \ref{fig:3}) simply by grouping multiple individual agents that had evolved to acquire chemotaxis. Despite having no information about other agents during the training phase, the results suggest that primitive behavioral principles like chemotaxis alone can enable swarm formation. This aligns with findings in biological systems where self-organization emerges from simple individual behaviors \cite{Camazine2020}.

Our results suggest that individual evolution does not necessarily lead to collective novel behaviors. Collective fitness does not directly correspond to individual fitness, and even when individual evolution stabilizes, collective diversity continues to change. We attribute the diversity of behavior emerged when individuals grouped together to due pheromones, a means of communication. This is in line with \cite{Olaf2016} that argues signal communication allows for diversity in collective behaviors. During periods of stable individual fitness, we observed a decrease in sensor-motor coupling $MI(\boldsymbol{I};\boldsymbol{O})$. This reduction in sensory sensitivity appears to lead to uniform movement patterns across the agent population. This phenomenon resembles the concept of collective intelligence as observed in natural systems, where group performance can diverge from individual capabilities \cite{bonabeau1999swarm}. We observed that mutual information $MI(\boldsymbol{I};\boldsymbol{O})$ is higher than the internal information content $H(\boldsymbol{O}|\boldsymbol{I})$. It demonstrates their behaviors emerge through interactions between the external environment and their internal states. We hypothesis that it is due to adaptability for individual fitness and robustness for collective behavior obtained through evolution are complementary \cite{Iizuka2004}. We leave the better way to balance adaptability and robustness to further studies.

The collective behavior at generation $100$ is particularly interesting, showing remarkable similarities to natural ant colonies. At this generation, the population differentiates into agents exhibiting local movement patterns and those performing global movements. Similar differentiation of roles has been observed in social insects \cite{robinson2009division} and \cite{gordon2016division}. Our analysis suggests that this emergence of two behavior patterns is supported by massive information flow from the environment. This external information appears to drive internal information flow, resulting in complex behavioral patterns.

The single agent's individual fitness---its ability to collect pheromones---stabilizes at a constant level. Despite this, during a certain evolutionary process, the collective fitness of the group declines at a constant rate. This decline indicates that the neural network possessed by individual agents is changing. This result suggests that when the network connection strength drifts randomly (likely within a neutral fitness landscape), the collective fitness, or the amount of pheromones collected—moves in a declining direction. This might imply that global movements that break free from localized motions may serve as attracting states in evolutionary dynamics.

In the population model of this study, each individual possesses an evolved neural network and makes decisions independently. The neural networks around generation $500$ show differentiation in population abilities, which leads to high group fitness. The dispersion of individual abilities has been proposed as an important condition for a population to become intelligent \cite{surowiecki2005wisdom}, and the population model around generation $500$ in this study best meets this condition.

Our findings indicate that collective AI can emerge through the differentiation of initially homogeneous agents, eliminating the necessity for pre-designed heterogeneity. This supports our ``Community First Hypothesis,'' which posits that homogeneous agents inherently possess the potential for heterogeneity, with society serving as the foundation for such agents. This is evidenced by the alignment of peak collective fitness with maximum variance in individual pheromone gain, while individual movement patterns exhibited minimal variation during this period. This suggests that the collective's success arises not from the superiority of individual agents but from the diversity among them. We highlight the significance of differentiation across multiple organizational levels, from individual agents to the collective, in fostering collective intelligence. These insights offer valuable guidance for the future architectural design of collective AI systems.

\section*{Acknowledgments}
This work was partially supported by Google Cloud and Google University funding (2022). It is also partially supported by Grant-in-Aids Kiban-A (JP21H04885) and Grant-in-Aids for JSPS Fellows (JP24KJ0753).

\newpage
\section*{Appendix 1: Time Series Analysis of Chemotaxis}
\begin{figure}[htbp]
\centering
\includegraphics[width=\columnwidth]{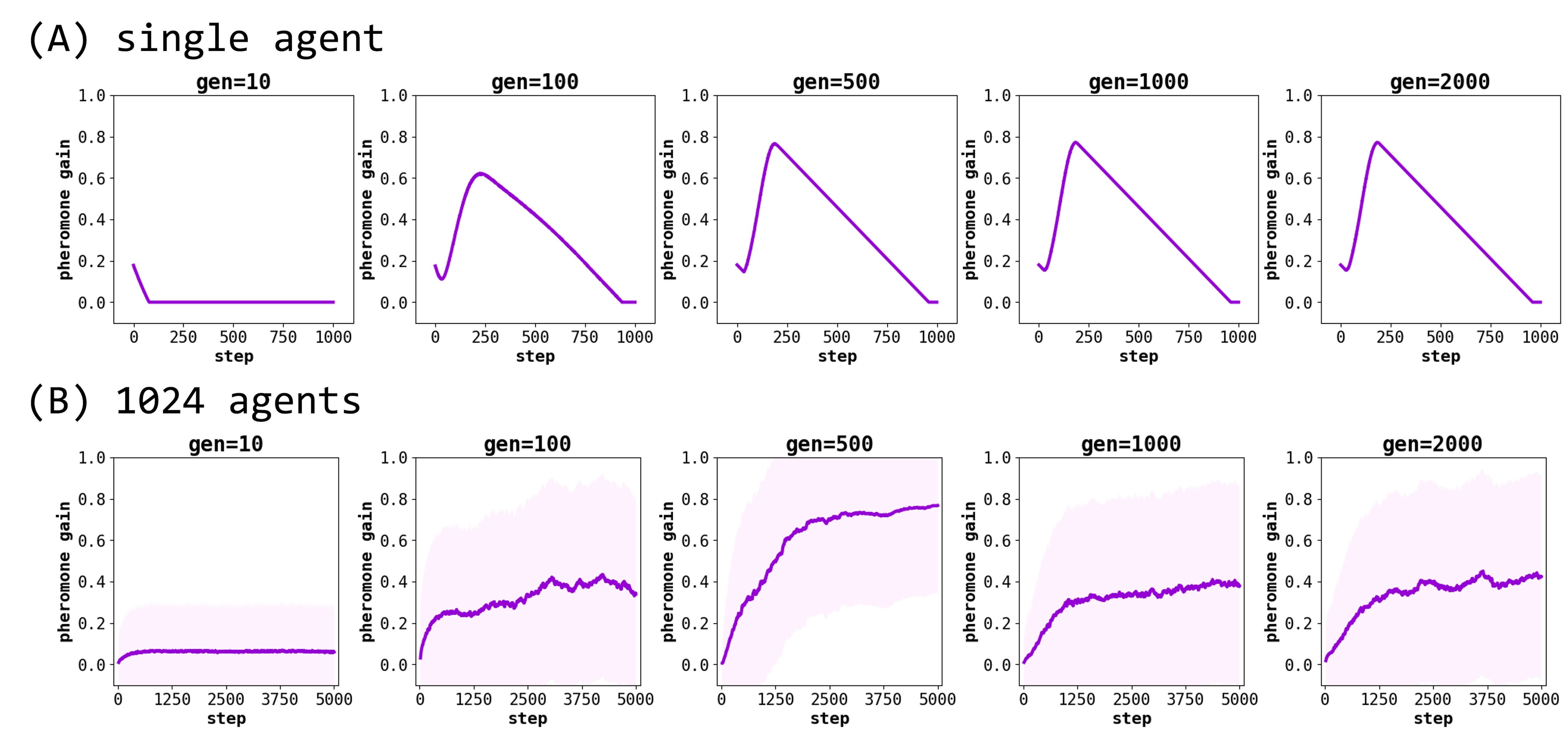}
\caption{Time series of pheromone gain at generations $10$, $100$, $500$, $1000$, and $2000$, corresponding to the pheromone patterns in Figure \ref{fig:3}.
\textbf{(A)} Single agent evolves over $1000$ steps to climb pheromone gradients. It learns to reach the peak of the evaporating pheromone hill as quickly as possible.
\textbf{(B)} Average pheromone gain over $5000$ steps for $1024$ agents. The middle generations show the highest gain levels, while later generations show decreased pheromone gain levels.}
\label{fig:a1}
\end{figure}

Time series data showing the evolution of pheromone gain in both single-agent and multi-agent scenarios (Figure \ref{fig:a1}), corresponding to Figure \ref{fig:3}. Single agents begin climbing pheromone gradients around generation $100$. The decrease in pheromone gain in the latter half is due to natural pheromone evaporation. The average pheromone gain by the collective peaks at generation $500$, then declines to levels similar to generation $100$.

\begin{figure}[htbp]
\centering
\includegraphics[width=8cm]{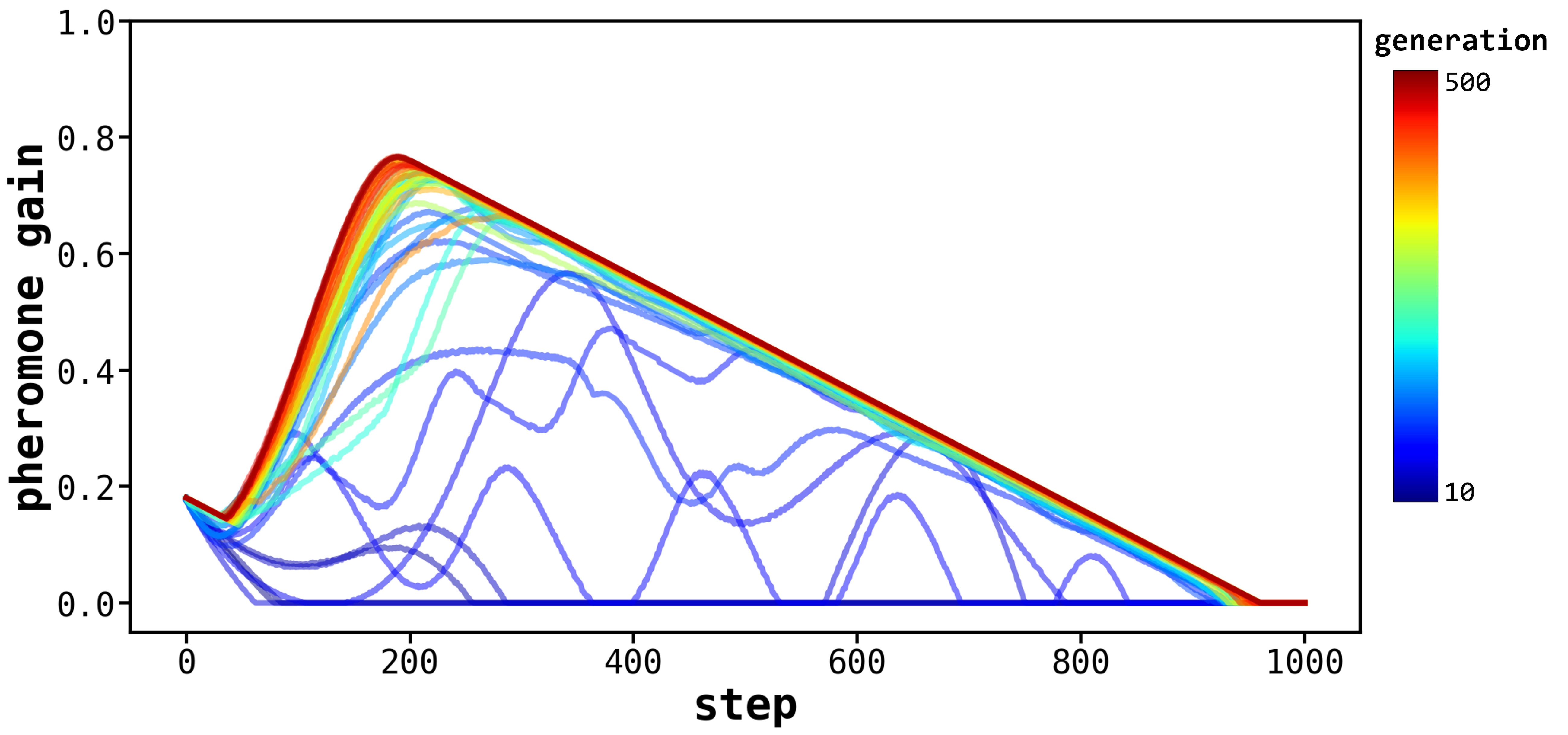}
\vspace{-0.1cm}
\caption{Evolution of single agent pheromone gain up to generation $500$. Chemotaxis is optimized so that the agent reaches the peak of the pheromone gradient in the shortest steps.}
\label{fig:a2}
\end{figure}

Time series analysis of single agent pheromone gain up to generation $500$ (Figure \ref{fig:a2}) reveals gradual chemotaxis development. Early in evolution, agents show varied responses to pheromone gradients, but eventually develop strategies to reach areas of highest pheromone concentration via optimal paths.

\vspace{-0.2cm}

\vspace{-0.5cm}

\newpage
\section*{Appendix 2: Dynamics of Context Neurons}
\begin{figure}[htbp]
\centering
\includegraphics[width=\columnwidth]{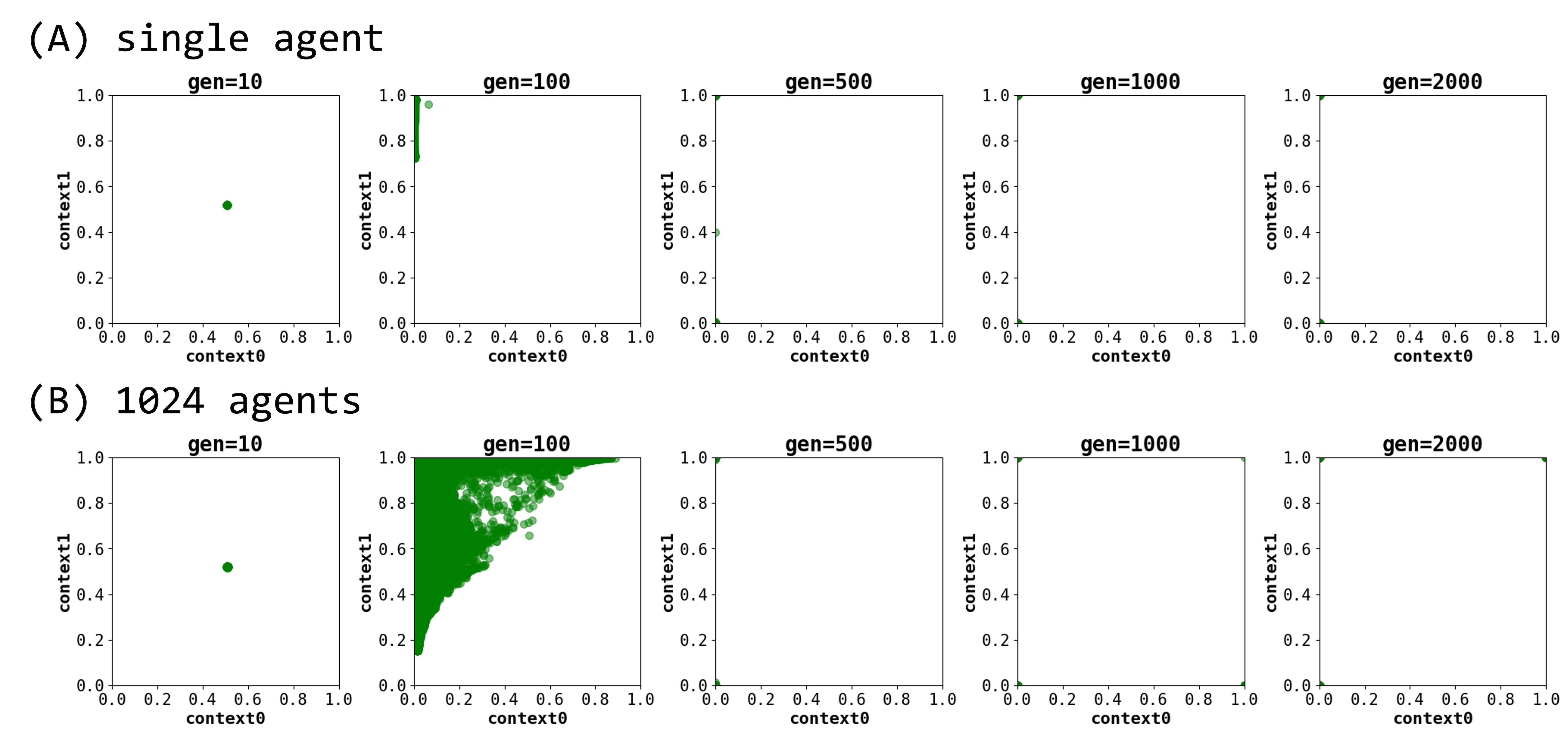}
\caption{Mapping of two context neurons' outputs from the agent's neural network, corresponding to the pheromone patterns in Figure \ref{fig:3}.
\textbf{(A)} Context neuron outputs over $1000$ steps for a single agent.
\textbf{(B)} Context neuron outputs sampled over $1000$ steps from one of $1024$ agents. The most dynamic responses are observed at generation $100$.}
\label{fig:a3}
\end{figure}

Mapping of context neurons in both single-agent and multi-agent scenarios (Figure \ref{fig:a3}), corresponding to the environments shown in Figure \ref{fig:3}. Context neurons exhibit their most dynamic behavior at generation $100$, leading to differentiation between local and global behaviors within the agent population. After generation $500$, both context neurons shift to discrete states, outputting only $0$ or $1$.

\newpage
\section*{Appendix 3: Distribution of Collective Behavior Data}
\begin{figure}[htbp]
\centering
\includegraphics[width=\linewidth]{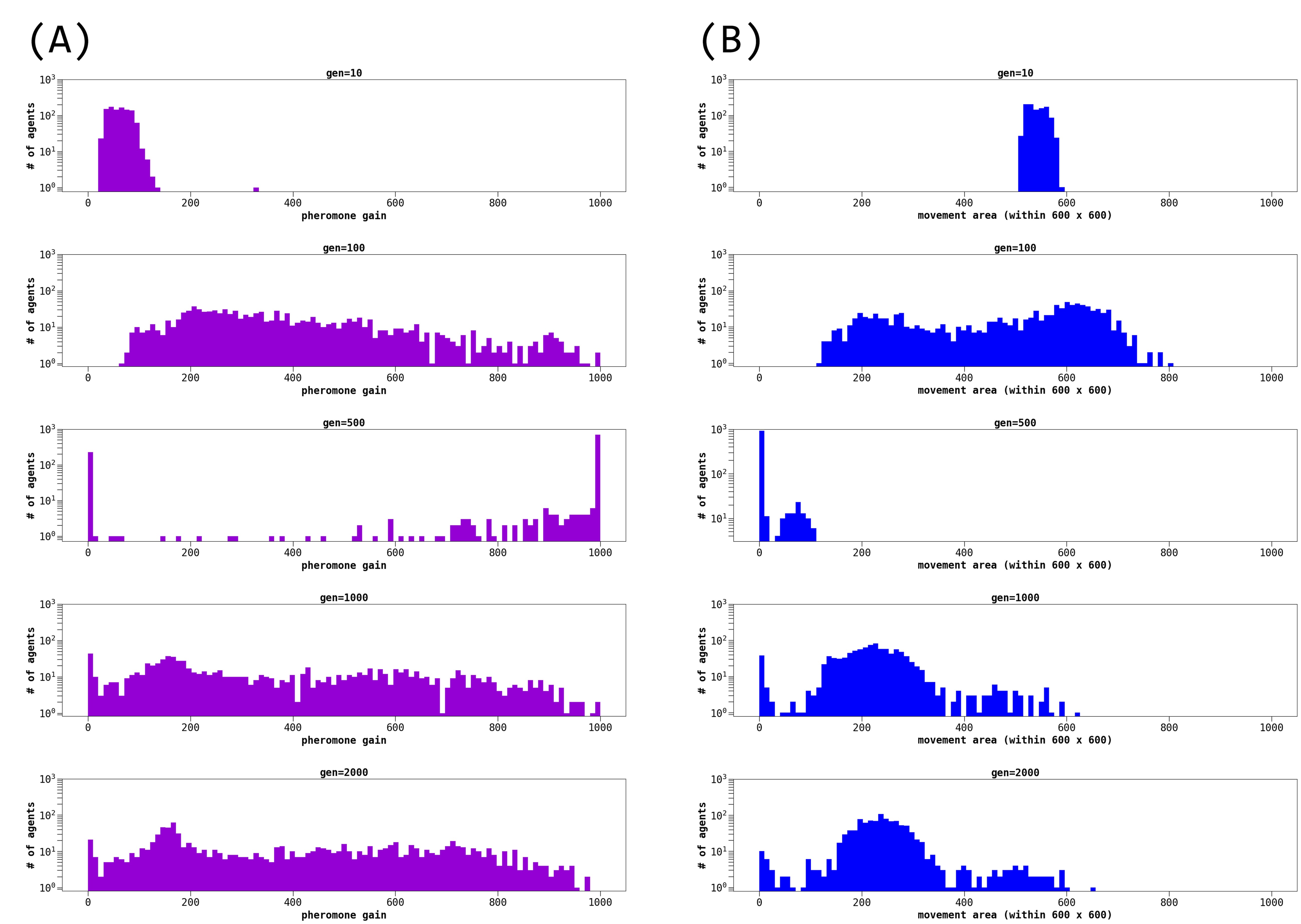}
\caption{Evolution of collective behavior distributions at generations $10$, $100$, $500$, $1000$, and $2000$. The standard deviations shown in Figure \ref{fig:4} are calculated from these distributions.
\textbf{(A)} Distribution of pheromone gain. At generation $500$, the distribution concentrates at both extremes, then spreads across the entire range in later generations.
\textbf{(B)} Distribution of movement area. At generation $500$, the movement area distribution becomes skewed toward the left side.}
\label{fig:a4}
\end{figure}

Distribution of pheromone gain and movement area for $1024$ agents (Figure \ref{fig:a4}), corresponding to Figures \ref{fig:4} (B) and (C). In early generations, all agents show low pheromone gain and localized movement patterns. At generation $100$, both distributions widely expand, and after generation $500$, completely stationary agents emerge. Generation $500$ notably shows differentiation between agents acquiring no pheromones and those achieving maximum gain, coinciding with peak collective fitness.

\newpage
\section*{Appendix 4: Results of Different Evolution Seed}
\begin{figure}[htbp]
\centering
\includegraphics[width=\linewidth]{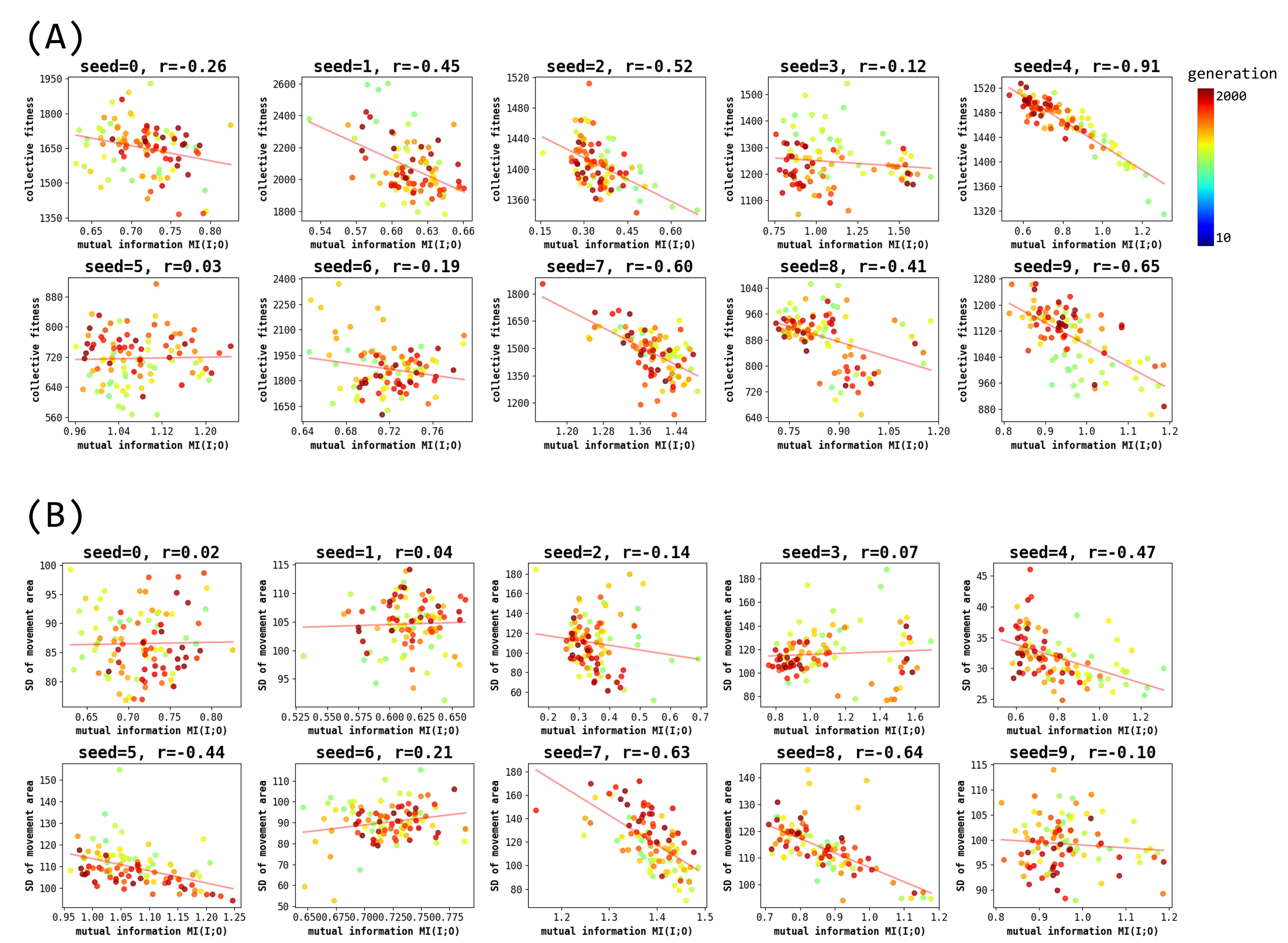}
\caption{Correlations between mutual information $MI(\boldsymbol{I};\boldsymbol{O})$ and corresponding metrics in later generations across $10$ different evolutionary seeds. Colors indicate generations.
\textbf{(A)} Negative correlation between mutual information and collective fitness.
\textbf{(B)} Negative correlation between mutual information and standard deviation of movement area.
These relationships suggest that when agents behave independently of environmental information, collective movement becomes more coordinated and collective fitness increases.}
\label{fig:a5}
\end{figure}

Analysis across $10$ different evolutionary seeds shows negative correlations in later generations between mutual information $MI(\boldsymbol{I};\boldsymbol{O})$ and both collective fitness and movement area standard deviation (Figure \ref{fig:a5}). Collective fitness shows stronger negative correlation, suggesting that, despite seed variation, higher collective fitness corresponds to weaker input-output relationships.

\newpage
\section*{Appendix 5: Rule-based Agent Without Internal States as a Criterion}
To verify that the agent has acquired chemotaxis, we created a rule-based agent. The body, sensors, and motors of the agent are the same as those of the neural network agent. The agent proceeds in the direction of the sensor position to which it responded most strongly among the sensors around its body. Specifically, for each sensor input, the motor output is as shown in Table \ref{tab:3}.

\begin{table}[htbp]
\small\sf\centering
\caption{Motor outputs of rule-based agent. This agent has designed chemotaxis behavior.\label{tab:3}}
\begin{tabular}{c|rr}
\toprule
\multirow{2}{*}{Most responsible sensor ID} & \multicolumn{2}{c}{Motor Outputs}\\
& Velocity & Angular velocity\\
\midrule
0 & 1.0 & 0.00\\
1 & 0.6 & 0.02\\
2 & 0.2 & 0.04\\
3 & 0.2 & -0.04\\
4 & 0.6 & -0.02\\
5 & 0.0 & 0.00\\
\bottomrule
\end{tabular}
\end{table}

The rule-based agent here has no internal states, only responses to inputs are implemented. With reference to such direct chemotaxis coupled to the environment, we examine the dynamics due to chemotaxis via internal states implemented by a neural network.

\begin{figure}[htb]
\centering
\includegraphics[width=4cm]{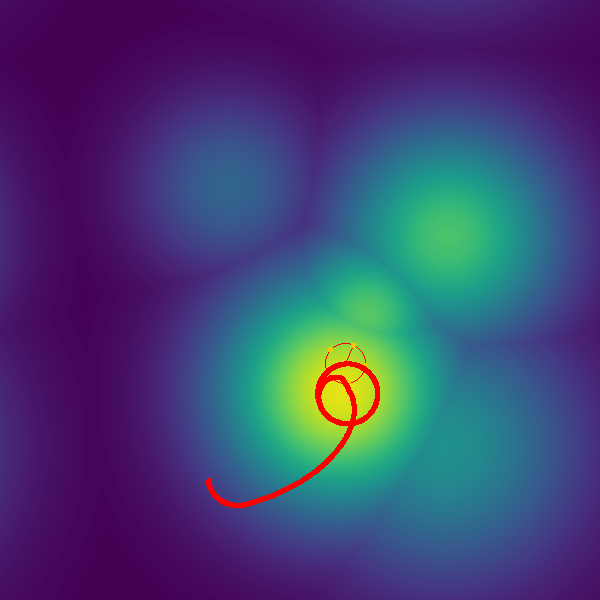}
\caption{Trajectory of rule-based agent. The agent exhibits chemotaxis in response to pheromone gradients.}
\label{fig:a6}
\end{figure}

The chemotaxis acquired by the agent through the evolution of the neural network was compared to the rule-based chemotaxis. The trajectory of the agent, without internal states and with rule-based chemotaxis, is shown in Figure \ref{fig:a6}. The trajectories of the agents are represented for $1000$ steps in the single-agent environment. The behavior of the agent climbing the pheromone gradient shown here is similar to the behavior acquired through evolution in Figure \ref{fig:3} (A).

\begin{figure}[htb]
\centering
\includegraphics[width=10cm]{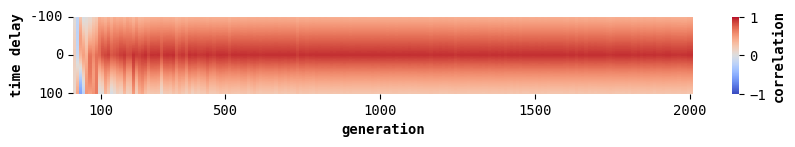}
\caption{Cross-correlation between trajectories of rule-based agent and evolved single agent. The similarity between trajectories was measured as both agents moved through the same pheromone environment. Colors indicate correlation strength. After generation $100$, the two behaviors show strong similarity.}
\label{fig:a7}
\end{figure}

Here, to measure the similarity of the behaviors, we measured the cross-correlation of the time series of behavioral outputs through $1000$ steps (Figure \ref{fig:a7}). This cross-correlation represents the degree to which the evolved chemotaxis is similar to the rule-based chemotaxis. In this Figure, the horizontal axis represents the number of steps, the vertical axis represents the time delay, and the colors represent the strength of the correlation. It can be seen that after approximately generation $100$, the behaviors are synchronized with almost no time delay. This means that the rule-based level of chemotaxis was acquired in around $100$ generations.

\begin{figure}[htb]
\centering
\includegraphics[width=\linewidth]{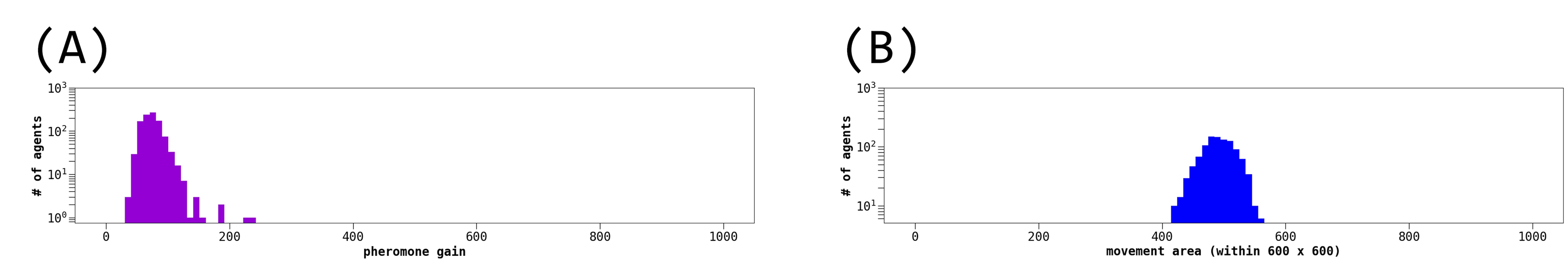}
\caption{Distribution of behaviors in multi-agent simulation with $1024$ rule-based agents, compared with Figure \ref{fig:a4}.
\textbf{(A)} Distribution of pheromone gain.
\textbf{(B)} Distribution of movement area.
Rule-based agents without internal states do not show behavioral differentiation.}
\label{fig:a8}
\end{figure}

Individual behavior was not differentiated in the population in the rule-based agent (Figure \ref{fig:a8}). These agents have no internal state and act only in response to inputs. This suggests that for individual behavior to be differentiated, it is necessary to have an internal state, such as a neural network.

%Bibliography
\bibliographystyle{unsrt}  
\bibliography{references}

\end{document}